\documentclass[aps,floats,prl,
showpacs, twocolumn]{revtex4}

\usepackage{amsfonts,amsmath} \usepackage{bm} \usepackage{dcolumn}
\usepackage{epsfig} \usepackage{latexsym}

\begin{document}

\title{Superdiffusion in the topological metal}

\author{Chushun Tian$^{1,2}$}

\affiliation{$^1$ Institute for Advanced Study, Tsinghua University, Beijing, 100084, China\\
$^2$ Institut f{\"u}r Theoretische Physik, Universit{\"a}t zu K{\"o}ln, K{\"o}ln, 50937, Germany}

\begin{abstract}
{\rm We develop a non-perturbative theory to study
large-scale quantum dynamics
of Dirac particles in disordered scalar potentials (the so-called ``topological metal'').
For general disorder strength and carrier doping,
we find that at large times, superdiffusion occurs.
I.e., the mean squared displacement grows as
$\sim t\ln t$. In the static limit, our analytical theory shows that
the conductance of a finite-size system obeys the scaling equation identical to
that found in previous numerical studies.
These results suggest that in the topological metal,
there exist some transparent channels -- where waves propagate ``freely'' --
dominating long-time transport of the system.
We discuss the ensuing consequence -- the transverse superdiffusion
in photonic materials -- that might
be within the current experimental reach.}
\end{abstract}


\maketitle

The past years have witnessed that graphene \cite{Geim09,DasSarma11}
and topological insulators \cite{Hasan10}
become an exciting frontier of
condensed matter physics.
A characteristic feature common to these novel materials is
the emergence of the massless Dirac particle.
For example, it appears on the surface of
three-dimensional strong
topological insulators such as
${\rm Bi}_2{\rm Se}_3$ and ${\rm Bi}_2{\rm Te}_3$ \cite{Hasan10,Kane07,Hasan08}.
This particle displays helical spin
(that accounts for the sublattice structure in graphene)
structure in momentum space
and is ``relativistic'': the
energy dispersion is linear.
On one hand, the discovery of the Dirac particle in these materials
is fostering conceptual developments in physics,
notably the Klein tunneling in condensed matter systems \cite{Geim06};
on the other hand, with many striking properties, the Dirac particle
may find considerable potential
applications in new electronic devices.

For experiments and practical applications,
the interplay between Dirac particles and disorders
is a key issue \cite{Liu11}. A
complete theory must include the description of quantum dynamics of
the Dirac particle in disordered environments. This is a subject
largely unexplored. Yet, there have been increasing evidences
indicating that rich dynamic phenomena might occur in this system. Indeed,
in the absence of disorders,
experimental and theoretical studies
(see Refs.~\cite{Peleg07,Zhu09} and references therein)
have shown that the Dirac particle has already exhibited
interesting dynamic behavior.
In addition, being in the same symmetry class notwithstanding,
a disordered Dirac particle system and a normal metal with disordered spin-orbit coupling
have transport properties of profound differences
\cite{Nomura07,Beenakker07,DasSarma09}.
It is thereby conceivable that in the presence of disordered potentials,
the interplay between wave interference and
the relativistic energy dispersion may lead to even
more interesting dynamic behavior and eventually,
to unusual electric and optical properties.

The purpose of this Letter aims at
a systematical and analytical investigation of
this subject.
Specifically, we shall focus on
the two-dimensional system with a single Dirac valley and subject to
scalar disordered potentials $V({\bf x})$, the so-called ``topological metal'' \cite{Nomura07}.
For this system, the quantum dynamics is characterized
by a two-component spinor, $\Psi$, obeying  (We set $\hbar=1$.)
\begin{equation}\label{eq:7}
    i\partial_t \Psi = \hat H \Psi, \quad \hat H\equiv -iv (\sigma^x \partial_x + \sigma^y \partial_y) +
V({\bf x}).
\end{equation}
Here, $v$ is the Fermi velocity and $\sigma^i, \,i=0,x,y,z$
stands for the Pauli matrices. The effective
time-reversal symmetry, i.e.,
$    \sigma^y \hat H^* \sigma^y =\hat H $,
brings the system to the sympletic symmetry class.
Below we develop a non-perturbative
theory to study quantum dynamics (\ref{eq:7}) at large scales.

Armed with this analytical theory,
we find that {\it for general disorder strength and carrier doping},
the quantum transport of this system
exhibits certain ``anomalies''. I.e.,
(i) at large times, quantum {\it superdiffusion} occurs: the mean squared displacement
-- characterizing the expansion of wave packets -- is given
by
\begin{equation}\label{eq:1}
\delta {\bf x}^2(t)
\stackrel{t\rightarrow \infty}{\longrightarrow} \frac{1}{4\pi^2 \rho_{\epsilon_F}}\,t\ln t,
\end{equation}
where $\rho_{\epsilon_F}$ is the density of states at
the energy $\epsilon_F$. It suggests that
such a topological metal behaves as a ``transition'' from Ohmic
($\delta {\bf x}^2\sim t$) to perfect ($\delta {\bf x}^2\sim t^2$) metals.
The $t\ln t$ behavior, though of pure quantum origin
(as we will see shortly),
is akin to a classical transport anomaly discovered long ago \cite{Bouchaud90}.
Experiences in studies of classical superdiffusion \cite{Bouchaud90} then
suggest that the finding of quantum superdiffusion
might have far-reaching impacts on many scientific branches including
statistical physics, condensed matter physics, nonlinear dynamics,
and pure mathematics. We notice that sufficiently
away from the Dirac point, this expression is
{\it universal}: it does not depend on disorders at all,
as $\rho_{\epsilon_F}$ converges to its clean limit.
In contrast, near the Dirac point, this expression is non-universal
because $\rho_{\epsilon_F}$
is governed by disorders.
(ii) In the static limit, our analytical theory shows that for a disordered sample of size $L$,
the conductance $g(L)$ follows the scaling equation,
\begin{equation}\label{eq:5}
    \frac{d\ln g}{d\ln L}=\frac{1}{\pi g},
\end{equation}
irrespective of $g$. This is consistent with numerical findings
\cite{DasSarma11,Nomura07,Beenakker07,DasSarma09}. Remarkably,
unlike a normal metal with disordered spin-orbit coupling, this system does not exhibit
two-dimensional Anderson transition. The nature of
anti-localization in topological metals is
currently under intense investigations \cite{Nomura07,Ryu07}.

\begin{figure}[h]
  \centering
\includegraphics[width=8cm]{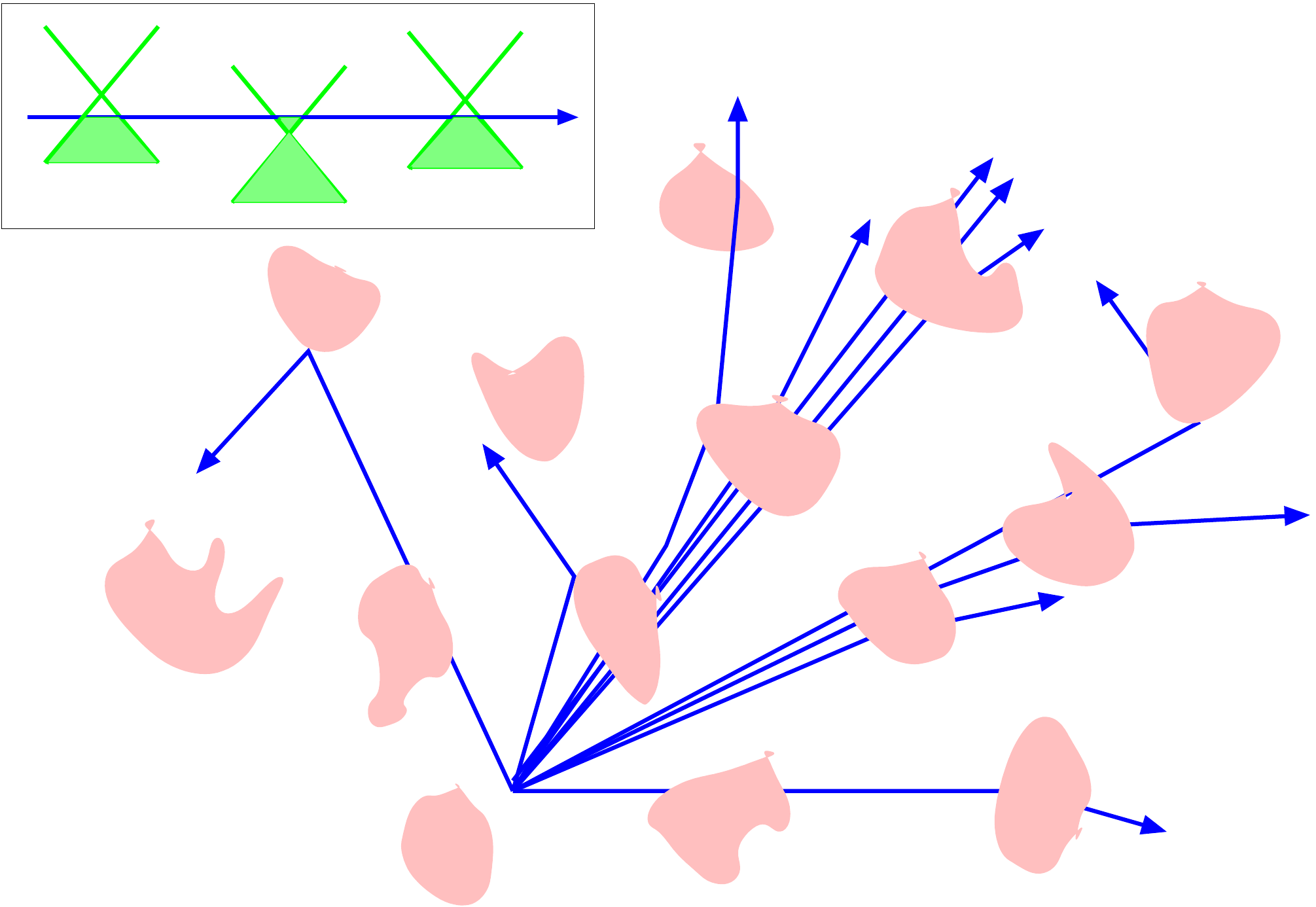}
\caption{Multiple scattering of a Dirac particle by disordered potentials (pink).
Inset: Klein tunneling.}
  \label{fig1}
\end{figure}

Let us first discuss a possible physical picture
underlying Eqs.~(\ref{eq:1}) and (\ref{eq:5}) and give their
``intuitive'' derivation. In doing so, we will see that (i) long-time transport
of the system is dominated by transparent channels where
waves propagate ``freely''; and
(ii) the full quantum diffusion coefficient (including multiple scattering effects)
displays logarithmic singularity.

We shall focus on the Dirac point ($\epsilon_F=0$).
In this case, the Klein tunneling \cite{Geim06} generally occurs (Fig.~\ref{fig1}, inset):
upon hitting a potential barrier, a particle may
be converted into a hole and subsequently undergoes perfect transmission;
similar scenario may happen to a hole.
On general grounds, we expect (Fig.~\ref{fig1}) that assisted by the
Klein tunneling, particles acquire a significant probability
of passing ``freely'' through disordered potentials before
deflecting from the incidence direction, leaving a
``free'' flight path (much) longer than the 
distance between two nearest disordered potentials. In the extreme case,
they keep moving along the incidence direction and 
pass freely through all the disordered potentials encountered, 
leaving an infinitely long free flight path (``transparent channels'').
Thus, multiple scattering leads to a length distribution
of the free flight path.

The diffusion coefficient
scales as $\sim v^2 \gamma^{-1}$, where
the scattering rate $\gamma$ 
is inversely proportional to the length 
of the free flight path defined above and thereby 
includes multiple scattering effects.
Averaging $v^2 \gamma^{-1}$ with respect to the
scattering rate distribution $P(\gamma)$ gives
the long-time asymptotic of the
diffusion coefficient,
$D(t^{-1}\rightarrow 0) \sim v^2 \int d
\gamma P(\gamma)\gamma^{-1} $.
Because of $P(0)\sim (\rho_{\epsilon_F} v^2)^{-1}$,
the integral over $\gamma$ suffers logarithmic divergence
and is dominated by $\gamma \sim 0$. That is, long-time
transport of the system is dominated by the (almost) 
transparent channels, i.e., $\gamma
\rightarrow 0$.
Since the free flight path cannot be
longer than $vt$ at finite time, i.e., $v\gamma^{-1}\leq vt$.
the scattering rate has a lower bound, $t^{-1}$.
(The upper bound is of minor importance as we are interested in the time-dependent
behavior.)
Taking this into account,
\begin{equation}\label{eq:2}
    D(t^{-1}\rightarrow 0) \sim v^2 P(0)\int_{t^{-1}} d
\gamma \gamma^{-1} \sim \ln t/\rho_{\epsilon_F}.
\end{equation}
Substituting it into the mean squared displacement, $\delta {\bf x}^2(t) \sim D(t^{-1}\rightarrow 0) t $, 
we reproduce Eq.~(\ref{eq:1}).
For a finite system, $t^{-1}$ scales as $\sim L^{-2}$. In combination with
the Einstein relation, Eq.~(\ref{eq:2}) gives the conductance $g(L)\sim e^2 \ln L$.
Therefore, we reproduce Eq.~(\ref{eq:5}) also.
Surprisingly, Eq.~(\ref{eq:2}) is {\it insensitive} to the details of
$P(\gamma)$. This universality
may reflect the topological nature of the system.

We are now ready to present the analytical theory
and some technical details of the proof. We begin with weakly disordered
scalar potentials that are Gaussian distributed,
$\overline{V({\bf x})}=0\,, \quad \overline{V({\bf x})V({\bf
    x}')}=\frac{v^2}{\pi g_0} \delta({\bf x}-{\bf x}')$.
Here, the dimensionless parameter $g_0$ characterizes
the disorder strength, and the overline stands for the disorder average.
To proceed further we introduce the two-particle correlation function,
$
\phi^\mu({\bf x}-{\bf x}',\omega)\equiv
  -\frac{1}{2\pi i}
{\rm tr}\overline{\sigma^\mu G^R_{\epsilon_F+\omega}({\bf x},{\bf x}')
G^A_{\epsilon_F}({\bf x}',{\bf x})},\, \mu=0,x,y
$, where the matrix
retarded (advanced) Green function obey
$(\epsilon-\hat H \pm i\delta)G^{R,A}_\epsilon
({\bf x},{\bf x}')=\delta ({\bf x}-{\bf x}')\, \sigma^0$. Notice that the disorder
average restores the translational invariance and correlation
functions thereby depend only on ${\bf x}-{\bf x}'$.
As such, we may pass to the Fourier
representation, $\phi^\mu ({\bf x}-{\bf x}',\omega) \rightarrow \phi^\mu({\bf q},\omega)
\equiv \phi^\mu(q)$ and likewise,
$\overline{G^{R,A}_{\epsilon}({\bf x},{\bf x}')}
\rightarrow \int d^2({\bf x}-{\bf x}') e^{-i{\bf p}\cdot
({\bf x}-{\bf x}')}\overline{G^{R,A}_{\epsilon}({\bf x},{\bf x}')}
\equiv {\cal G}_{\epsilon}^{R,A}({\bf p}) $. Here, ${\cal G}_{\epsilon}^{R,A}$ satisfies
$(\epsilon - v\sigma^\alpha p_\alpha -\Sigma^{R,A}_\epsilon({\bf p})){\cal G}_{\epsilon}^{R,A}({\bf p})
=\sigma^0$ \cite{note1}. For weak disorders, one may invoke
the self-consistent Bonn approximation (SCBA) to calculate the
self-energy, $\Sigma^{R,A}_\epsilon$, obtaining
$\Sigma^{R,A}_\epsilon=\frac{v^2}{2\pi g_0}\sigma^0 {\rm tr} {\cal G}^{R,A}$.
(Notice that the trace operation includes the momentum index.)
The density of states $\rho_{\epsilon}
\equiv -\frac{1}{\pi} {\rm tr} {\rm Im}\, {\cal G}_{\epsilon}^R$
is related to the imaginary
part of the self-energy via $\rho_{\epsilon}=-\frac{g_0}{v^2}{\rm tr} {\rm Im}\, \Sigma_{\epsilon}^R$.
Importantly, the SCBA gives an exponentially small
energy scale $\sim e^{-2\pi^2 g_0}$ \cite{Mirlin09}: for energies far below
this scale (the Dirac regime),
the density of states saturates, $\rho_\epsilon\sim e^{-2\pi^2 g_0}$,
while for energies far above it, $\rho_\epsilon$ converges to its clean limit
$ \sim |\epsilon|$.

Then, by adaption of the method of Refs.~\cite{Woelfle80,Nieh98},
it can be shown that in the hydrodynamic limit,
$q\rightarrow 0$,
\begin{equation}\label{eq:53}
    \omega \phi^0 (q)-q_\alpha \phi^\alpha (q)=-\rho_{\epsilon_F}.
\end{equation}
This is the macroscopic continuity equation
reflecting the particle
conservation law, where $\phi^0(q)$ is the
macroscopic density and $(\phi^{x}(q),\,
\phi^{y}(q))$ the macroscopic current.
The latter follows a Fick-like law, read
\begin{equation}\label{eq:59}
    \phi^\alpha(q) =
    -i D(q) q^\alpha \phi^0(q)
\end{equation}
in the Fourier representation. Here, $D(q)\equiv v^2/(-i\omega+\gamma(q))$
is full quantum diffusion coefficient, where $\gamma(q)$ is
the so-called ``current relaxation kernel'' given by
\begin{eqnarray}\label{eq:60}
    \gamma(q) = \frac{1}{\tau} + \frac{1}{2\pi \rho_{\epsilon_F}}
    {\rm tr}
    \left\{\hat q_\alpha\sigma^\alpha\Delta {\cal G}_q
  U_q\Delta {\cal G}_{q}\hat q_\beta\sigma^\beta\right\}.
\end{eqnarray}
Here,
$\Delta {\cal G}_q({\bf p}) \equiv {\cal G}^R_{\epsilon_F+\omega}({\bf p}+{\bf q})-
{\cal G}^A_{\epsilon_F}({\bf p})$,
the elastic scattering rate $\frac{1}{\tau}= -2{\rm tr}{\rm Im}\,\Sigma^R=v^2\rho_{\epsilon_F}/g_0
$, and
$U_q({\bf p},{\bf p}')$ is the two-particle irreducible vertex function.
This equation, together with the expression of $D(q)$,
builds up a bridge between the macroscopic hydrodynamics
and the microscopic quantum dynamics.
It is evident from Eq.~(\ref{eq:60}) that
localization effects -- encoded by $U_q({\bf p},{\bf p}')$ -- introduce renormalization of the
elastic scattering rate. In the absence of
quantum interference, Eq.~(\ref{eq:60})
leads to a Boltzmann diffusion coefficient
$
D \equiv v^2 \tau$
(with the classical transport mean free time twice
larger than the elastic scattering time).
Eqs.~(\ref{eq:53})-(\ref{eq:60})
constitute the framework of subsequent analysis.

The closed set of macroscopic equations (\ref{eq:53}) and (\ref{eq:59})
gives an exact relation between
the optical conductivity, $\sigma(q)$, and the full quantum
diffusion coefficient, $D(q)$. To this end,
let us introduce the density response function defined as
$\chi({\bf x}-{\bf x}',t-t')\equiv i
\overline{\langle[\Psi^\dagger({\bf x},t)\Psi({\bf x},t),
\Psi^\dagger({\bf x}',t')\Psi({\bf x}',t')]\rangle}$, with $\langle\cdot\rangle$
the ground state average. Upon passing to the Fourier
representation, $\chi({\bf x}-{\bf x}',t-t')\rightarrow \chi(q)$, one may
follow the derivation of Ref.~\cite{Woelfle80} to show
$\chi(q)\stackrel{q\rightarrow 0}{=}\omega\phi^0(q)+\rho_{\epsilon_F}$.
In combination with the solution to Eqs.~(\ref{eq:53}) and (\ref{eq:59}),
it gives
$\chi(q)=\rho_{\epsilon_F} \frac{D(q){\bf q}^2}{-i\omega + D(q){\bf q}^2}$.
As a result,
\begin{equation}\label{eq:64}
    \sigma(q) \equiv -e^2 \lim_{{\bf q}\rightarrow 0} \left(
    \frac{i\omega}{{\bf q}^2} \chi(q)\right)=e^2 \rho_{\epsilon_F} D(q).
\end{equation}
Thus, we justify the general Einstein relation for disordered Dirac particles.

The system's behavior at large scales is determined by
the current relaxation kernel, $\gamma(q)$.
Therefore, we proceed to
perform non-perturbative analysis of this kernel
dominated by the infrared divergences of the two-particle irreducible vertex
function $U_q({\bf p},{\bf p}')$. Generally, these divergences include two types:
the singlet component of diffuson and cooperon \cite{Hikami80}.
It can be shown that the effective time-reversal symmetry
leads to a mathematically rigorous theorem:
\\
\\
\noindent {\it The current relaxation kernel $\gamma(q)$ suffers no
diffuson-type infrared divergences.}
\\
\\
It generalizes the Vollhardt-W{\"o}lfle theorem
discovered for spinless electron systems \cite{Woelfle80}.

Thanks to this theorem, to calculate the two-particle
irreducible vertex, $U_q({\bf p},{\bf p}')$, we need to consider
the diagrams composed of (the singlet component of) cooperon.
For large frequencies, the dominant contribution to $U_q({\bf p},{\bf p}')$
consists of single cooperon, giving
a quantum correction to the (bare) elastic scattering rate
$ \sim -\frac{1}{\pi \rho_{\epsilon_F}\tau}
\int\frac{d^2{\bf k}}{(2\pi)^2}
(-i\omega +D {\bf k}^2)^{-1}$ order of ${\cal O}(g_0^{-1})$.
This is the well-known
weak anti-localization, with the negative sign
reflecting the destructive interference in the presence of the
$\pi$ Berry phase \cite{Nomura07,Beenakker07,Hikami80,Ando98}.
It should be noted, however, that even for weak disorders, this quantum
correction is divergent in the low-frequency limit $\omega\rightarrow 0$.
In this case the elastic scattering rate (and thus
the Boltzmann diffusion coefficient) undergoes strong renormalization and as such,
the perturbative expansion in $1/g_0$ breaks down. To go beyond the
perturbation,
we must take into account the full
two-particle irreducible vertex for Eq.~(\ref{eq:60}).
As a matter of the reciprocity principle, $U_q({\bf p},{\bf p}')\sim \phi^0({\bf p}+{\bf p}',\omega)
\sim [-i\omega + D(\omega)({\bf p}+{\bf p}')^2]^{-1}$,
where $D(\omega) \equiv D({\bf q}=0,\omega)$. This amounts to
the replacement of the Boltzmann diffusion coefficient in
the weak anti-localization correction by the dynamical
diffusion coefficient $D(\omega)$. In doing so, we obtain a
self-consistent equation of the dynamical
diffusion coefficient,
\begin{equation}\label{eq:3}
\frac{D}{D(\omega)}=1 -\frac{1}{\pi \rho_{\epsilon_F}}
\int\!\! \frac{d^2 {\bf k}}{(2\pi)^2}\frac{1}{-i\omega + D(\omega) {\bf k}^2},
\end{equation}
that is expected to hold for arbitrarily low frequencies.
Eq.~(\ref{eq:3}) implies the validity of one-parameter scaling in
topological metals, consistent with non-perturbative
studies based on field theories \cite{Ryu07}.

Let us discuss below the physical consequences of
the general theory described by Eqs.~(\ref{eq:53})-(\ref{eq:3}).

\noindent {\it a) Superdiffusion.}---We prepare a wave packet
with the energy near $\epsilon_F\, (>0)$ and let it evolve.  Its expansion
is characterized by the full time-dependence of
the mean squared displacement
given by $\delta {\bf x}^2(t)=\int \frac{d\omega}{2\pi} (1-e^{-i\omega t})D(\omega)/\omega^2$.
For low frequencies, $\omega \tau\ll e^{-4\pi^2 g_0}$, Eq.~(\ref{eq:3})
may be solved analytically, giving $D(\omega)/D \approx
\frac{1}{4\pi^2 g_0}\ln \frac{1}{-i\omega \tau}$.
Substituting this result into the
expression of $\delta {\bf x}^2(t)$, we obtain
Eq.~(\ref{eq:1}) for $t\gg \tau e^{4\pi^2 g_0}$.
Noticing that $i\omega \sim t^{-1}$, we have
$D(t^{-1})/D \approx
\frac{1}{4\pi^2 g_0}\ln \frac{t}{\tau}$ namely Eq.~(\ref{eq:2}). It is thereby justified that
at large times, localization effects lead to
strong renormalization of the elastic scattering rate.

\noindent {\it b) Optical conductivity.}---According to the Einstein relation (\ref{eq:64}),
for low frequencies, $\omega \tau \ll e^{-4\pi^2g_0}$,
the optical conductivity reads $\sigma({\bf q}=0, \omega)
\approx (\frac{e}{2\pi})^2 \ln\frac{1}{-i\omega\tau}$ that strikingly, is universal:
it depends on the elastic scattering rate logarithmically.

\noindent {\it c) Static conductance of a finite system.}---We have focused on the dynamic
property of a bulk (infinite extended) system so far. Now, we switch to the
static conductance of a finite system of size $L$. This subject is
currently under intense investigations \cite{DasSarma11,Beenakker07,Nomura07,DasSarma09}.
In the static limit, Eq.~(\ref{eq:3})
is simplified to
\begin{equation}\label{eq:6}
\frac{D}{D(0)}\left[1+\frac{1}{
\pi
g_0}\int_{\frac{1}{L}<|{\bf k}|<\frac{1}{v\tau}} \frac{d^2 {\bf k}}{(2\pi)^2}\frac{1}{{\bf k}^2}\right]=1
\end{equation}
thanks to $\frac{-i\omega}{D(\omega)}\stackrel{\omega\rightarrow 0}
{\longrightarrow} 0$. This equation gives the static diffusion coefficient
$D(0)=D[1+\frac{1}{2\pi^2
g_0}\ln \frac{L}{v \tau}]$.

To translate $D(0)$ to a
static size-dependent conductance (identical to the conductivity in
two dimensions), $g(L)$, we apply
a static, infinitesimally small electric field $E$ to the system
say, along the $x$-direction. In response to the electric potential,
$(L-x)eE$, a density profile
is established across the sample, read $-eE\int_0^L dx' \chi(x-x',0)(L-x')=eE\rho_{\epsilon_F}(x-L)$.
In obtaining this result, we have used the fact that
in the real space, the static
density response function is
$\chi(x-x',\omega=0)=\rho_{\epsilon_F}\delta(x-x')$.
Results from this density profile a uniform diffusive current,
$-e^2 \rho_{\epsilon_F} D(0) E$, that
is balanced by a macroscopic electric current, $g(L) E$.
It then follows
$g(L)=e^2 \rho_{\epsilon_F} D(0)$, giving Eq.~(\ref{eq:5}) (where
the conductance is in unit of $e^2/h$.)
For large samples, $L/v\tau \gg e^{2\pi^2 g_0}$, the
conductance $
g(L)=\frac{e^2}{2\pi^2} \ln \frac{L}{v\tau}$ that depends logarithmically on the elastic
scattering rate and has a {\it universal} prefactor.

\noindent {\it d) General disordered potentials.}---Qualitatively,
the macroscopic equations (\ref{eq:53}), (\ref{eq:59}) and (\ref{eq:64})
are the results of particle conservation, while Eq.~(\ref{eq:3}) reflects
the nature of one-parameter scaling and one-loop self-consistency.
As such, we expect them to be universal for a
large class disordered potential. Specifically, the disorder potentials
-- remaining of the scalar type --
may be strong, i.e., $g_0\lesssim 1$, and (or) exhibits long-ranged
Gaussian correlations.
These dramatic modifications notwithstanding, they merely affect the microscopic parameters
namely $\rho_{\epsilon_F}$ and $D$ in Eqs.~(\ref{eq:53}), (\ref{eq:59}),
(\ref{eq:64}) and (\ref{eq:3}). Then, we may repeat the discussions of
$a)-c)$, obtaining Eqs.~(\ref{eq:1}) and (\ref{eq:5}).
The latter is consistent with numerical findings
\cite{DasSarma11,Beenakker07,Nomura07,DasSarma09}, despite that
the method here
differs essentially
from the routine one based on the Kubo formula \cite{Nomura07}.

\begin{figure}[h]
  \centering
\includegraphics[width=8cm]{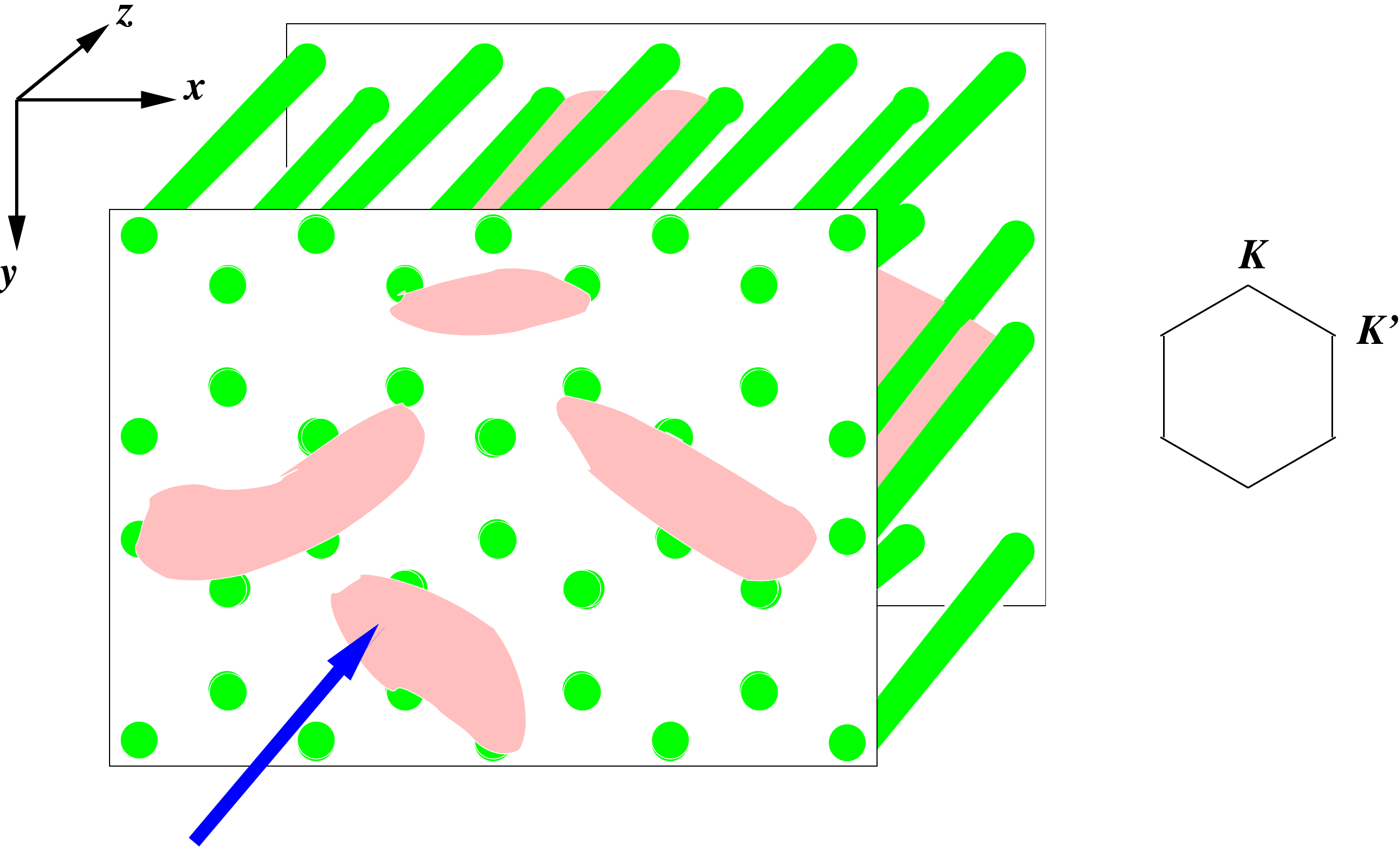}
\caption{Transverse superdiffusion. Left: a laser beam (blue) is launched into a two-dimensional
disordered photonic crystal. The latter is uniform in the longitudinal ($z$)
direction while in the transverse ($x$-$y$) plane,
it displays a periodic background of
honeycomb structure (green) with smooth random dielectric fluctuations
(pink). At a sufficiently large distance $z$,
the broadening of the probe beam scales as $\sim z\ln z$.
Right: the first Brillouin zone of the
honeycomb lattice.}
  \label{fig}
\end{figure}

In realistic electronic materials,
electron-electron interaction interplays strongly with wave interference effects
(as observed in experiments on topological insulators \cite{Liu11}).
Therefore, we discuss here the possibility of observing superdiffusion
in photonic materials that are free of such an interplay
(Fig.~\ref{fig}, left).
We adopt the method that leads to
the experimental discovery of the so-called
transverse localization \cite{Segev07}. In the present context,
one may invoke the optical induction technique
to fabricate a two-dimensional photonic crystal of thickness $z$
that displays the honeycomb structure in the transverse
($x$-$y$) plane while is uniform in the longitudinal ($z$) direction.
As such, the photonic band structure exhibits
two non-equivalent Dirac valleys, around $K$ and $K'$ (Fig.~\ref{fig}, right).
One may
further introduce random fluctuations of refractive-index
in the transverse plane that are smooth on a scale much larger
than the lattice constant. Then, a monochromatic laser beam
is launched into this disordered photonic crystal and
undergoes diffraction broadening. At a distance $z$ from the input plane,
the intensity profile of the probe beam is measured.

The probe beam excites only the Bloch modes in certain Dirac valley,
and has an initial width much larger than the scale
over which the refractive-index fluctuates. Its
propagation in the disordered photonic crystal
is described by the (rescaled) paraxial Hemlholtz equation \cite{Segev07},
$i\partial_z A=[-\frac{1}{2}(\partial_x^2+\partial_y^2) - n(x,y)]A$,
where $A(x,y,z)$ is the envelop of the electric field smoothly varying in the
longitudinal direction. Being static (having no-time dependence) notwithstanding,
this equation possesses a perfect analogy to
the Schr{\"o}dinger equation, with $z$ playing the role of ``time'' and
the (negative of) total refractive-index,
$- n(x,y)$,
of ``potential''. As a result, the diffraction broadening of the probe beam,
$A(x,y,z)$, mimics quantum wave function at time $z$. Because the ``potential''
is smooth enough, the inter-valley scattering is suppressed.
Effectively, this ``quantum dynamics'' is reduced to that described by Eq.~(\ref{eq:7}),
with the spin degree of freedom accounts for the sublattice structure.
Thus, we expect that the broadening of the probe beam (in the transverse plane),
$\int\!\!\int dxdy (x^2+y^2) |A(x,y,z)|^2$,
scales as $\sim z\ln z$ for sufficiently large thickness $z$. We term this phenomenon
{\it transverse superdiffusion}.

In summary, we develop a non-perturbative
theory of large-scale quantum dynamics of
Dirac particles in disordered environments. We show that
for general disorder strength and carrier doping, (i)
a bulk topological metal exhibits superdiffusion (\ref{eq:1}) at large times;
and (ii) the static conductance of a finite-size system
follows the scaling equation (\ref{eq:5}), consistent
with the numerical results \cite{DasSarma11,Nomura07,Beenakker07,DasSarma09}.
We discuss the ensuing consequence -- the transverse superdiffusion
in two-dimensional photonic materials -- that might
be within the current experimental reach.

I am deeply grateful to A. Altland,
T. Nattermann, H. T. Nieh,
Y. Y. Wang, Z. Y. Weng, S. C. Zhang, and Z. Q. Zhang
for valuable discussions.
Work supported by the NSFC (No. 11174174) and by the
Independent Scientific Research Program and the
Initiative Scientific Research Program of Tsinghua University.


\begin{thebibliography}{}
\bibitem{note1} Throughout this Letter the Greek indices $\alpha,\beta$ run over $x,y$ and
the Einstein convention is implied.
\end{thebibliography}
\end{document}